\documentclass[aps,twocolumn,noshowpacs,superscriptaddress,floatfix]{revtex4-1}

\usepackage[utf8]{inputenc}
\usepackage{braket}
\usepackage{amsmath,amssymb}
\usepackage[justification=justified]{caption}
\usepackage{qcircuit}
\usepackage{xcolor}
\usepackage{verbatim}
\usepackage[normalem]{ulem}
\usepackage[rightcaption]{sidecap}
\usepackage{subfig}
\usepackage{comment}

\usepackage[section]{algorithm} 
\usepackage{amsthm,algorithmic,epsfig,tikz}

\renewcommand{\part}[2]{\frac{\partial #1}{\partial #2}}

\usepackage{float}
\usepackage{multirow}
\usepackage[hidelinks]{hyperref}

\begin{document}

\title{Comparing Classical-Quantum Portfolio Optimization with Enhanced Constraints}

\author{Salvatore Certo}
\email{scerto@deloitte.com}
\affiliation{Deloitte Consulting, LLP}

\author{Anh Dung Pham}
\affiliation{Deloitte Consulting, LLP}

\author{Daniel Beaulieu}
\affiliation{Deloitte Consulting, LLP}

\date{\today}

\begin{abstract} 
One of the problems frequently mentioned as a candidate for quantum advantage is that of selecting a portfolio of financial assets to maximize returns while minimizing risk. In this paper we formulate several real-world constraints for use in a Quantum Annealer (QA), extending the scenarios in which the algorithm can be implemented.  Specifically, we show how to add fundamental analysis to the portfolio optimization problem, adding in asset-specific and global constraints based on chosen balance sheet metrics.  We also expand on previous work in improving the constraint to enforce investment bands in sectors and limiting the number of assets to invest in, creating a robust and flexible solution amenable to QA.

Importantly, we analyze the current state-of-the-art algorithms for solving such a problem using D-Wave's Quantum Processor and compare the quality of the solutions obtained to commercially-available optimization software.  We explore a variety of traditional and new constraints that make the problem computationally harder to solve and show that even with these additional constraints, classical algorithms outperform current hybrid solutions in the static portfolio optimization model. 

\end{abstract} 

\maketitle

\section{Introduction}

While quantum computers are theorized to solve problems currently intractable on classical computers, the current devices are noisy, small, and are limited with the types of algorithms feasible in today's devices.  While gate-based machines in the Noisy Intermediate Scale Quantum (NISQ) era struggle to find appropriate feasible applications, quantum annealers have less constraints and appear to be the most promising in near-term industrial implementations.

Quantum annealing follows the adiabatic theorem, which states that if the time evolution is long enough, a quantum system will stay in the ground state of its Hamiltonian.  For a quantum system starting in a ground state of a hamiltonian $H_b$ and evolving to the ground state of the problem hamiltonian $H_p$, the system can be described at time $t$ out of the total run time $T$ as:

\begin{equation}
\mathcal{H}(t) = (1-u(t))\mathcal{H}_b + u(t)\mathcal{H}_p
\end{equation}

where $u(t) = t/T$.  Quantum annealing allows for interesting problems to be solved by formulating the problem into the Hamiltonian $H_p$.  At the end of the annealing process, the system should be in the ground state and the solution to our problem found.

Combinatorial optimization problems are perfect candidates for this annealing process, as the problem Hamiltonian is easy to construct and is diagonal in the computational basis, thereby having the ground state being the exact readout of the qubits.  A well-known combinatorial optimization problem is that of portfolio optimization, where the task is to choose the assets in the right proportion as to maximize return while minimizing risk.  

This is an especially relevant problem not only for financial institutions, but for a variety of industries as non-financial assets can also be used to construct the problem \cite{bouland2020prospects}\cite{herman2022survey}\cite{mugel2020use}.  It is also easy to formulate, the necessary data is readily available, and the solutions are particularly significant for industrial applications.  The problem follows that of Markowitz's Modern Portfolio Theory \cite{markowitz} that the optimal portfolio is constructed with assets that maximize return for a given level of risk.  

This problem can be mapped to a Quadratic Unconstrained Binary Optimization (QUBO) problem defined by the problem hamiltonian:

\begin{equation}
\mathcal{H}_p = -\mu^Tw + q*w^T\Sigma w + k(\sum_{n}{w_n} - 1)^2
\end{equation}

where $\mu$ is the vector of logarithmic returns, $w$ is a vector of $n$ assets, where $w_n \in [0,1]$ the fraction of the total budget invested in each asset $n$, $\Sigma$ is the matrix specifying the covariance of assets, and $q$ is the given level of risk appetite for the investor.  Typically, one can adjust $q$ to map the target level of risk to the problem.  

As QA require the cost function to be unconstrained, we can add soft constraints with appropriate Lagrange Multipliers to enforce conditions.  For instance, we will want the total available budget to be invested, and so we can add to $\mathcal{H}_p$ the budget constraint $k(\sum_{n}{w_n} - 1)^2$ where we set $k$ to be sufficiently large enough to ensure the constraint is satisfied.

Portfolio optimization models have been implemented in various forms on QA \cite{cohen2020portfolio}\cite{cohen2020portfolio_60}.  Some simple implementations include binary decision variables for each asset specifying whether or not to include them in the portfolio, while others go a step beyond and create variables for the exact number of shares or proportion of the asset to purchase given the available budget.  In this paper we extend the latter model in several ways.  First we introduce slack variables as an alternative to enforce minimum and maximum investments in particular sectors, showing that this formulation has several benefits that make it advantageous compared to other methods.  We then show how we can construct global and local constraints based on quantitative features like balance sheet metrics to construct portfolios that not only maximize return but also have the desired characteristics found in fundamental analysis of stocks.  

As current QA do not have the number of qubits nor the required connectivity between them to implement large-scale models directly on annealers, we explore the use of D-Wave's hybrid models, specifically for implementing the QUBO representation of the model in their hybrid Binary Quadratic Model (BQM) as well as using integer variables and explicit constraints in their hybrid Constrained Quadratic Model (CQM).  We note that the QUBO formulations we present will be especially advantageous once the annealing hardware matures and the number of qubits increase, but also show the benefit of hybrid solvers in the current era.  We therefore provide direct comparisons between both implementations and finally compare the results against a leading optimization software package CPLEX.

\section{QUBO Model and Constraints}

First we should note that in the model from Equation 2, instead of $w_n$ being the weight invested in asset $n$, it can instead be formulated to include the current stock price with the goal being to specify how many shares of each asset we wish to hold, considering our total budget $B$.  This would transform our cost function to:

\begin{equation}
\begin{aligned}
& \mathcal{H}_p = -\mu^Tw + q*w^T\Sigma w + k(\sum_{n}{s_np_n} - B)^2 \\
& w = \frac{s_np_n}{B}
\end{aligned}
\end{equation}

where $s_n$ is the number of shares invested in asset $n$, $p_n$ the current asset price, and B the dollar value of our available funds.

This formulation ensures that we are only investing in bundles (or even lots) and not fractional shares.  While this is important, in reality only small investors would need to ensure that the proportion of funds invested in each asset is achievable in whole shares, as large portfolios can always be constructed to an exact number of whole shares based on the weighted proportion of that asset to the portfolio.  In any case, the rest of this paper can be reformulated for whole shares or round lots with a few substitutions of variables.

In the subsequent sections we provide formulations for common real-world constraints for the problem.  These constraints, once appropriated mapped for use in a QUBO problem, will be added to Equation 2 with penalty coefficients to enforce the condition as necessary.

\subsection{Binary to Integer Mapping}

As our goal is to construct the cost function with binary only integers as to be amenable to QA, we need a way to convert our binary variables to integer ones.  The easiest and most straightforward way to use a binary encoding \cite{hybrid_holding} and a specified bit depth $N_d$ and degree of precision $P$ in our weights:

\begin{equation}
w_n = \frac{1}{P} \sum_{d=0}^{N_d-1}2^d x_{n,d}
\end{equation}

where $x_{n,d}$ is a binary variable for each asset and bit.  Here $P$ will be the level of precision we want in our weights of assets to invest in.  A $P$ of 100 will have each asset being invested in proportions of 1\%, while a $P$ of 10,000 will allow us to have the smallest proportional weights of 1 basis point.  Based on our level of chosen precision, we can then choose $N_d$ such that $2^{N_d} > P$.

A common condition in real life scenarios is to ensure there is a sufficient level of diversification in the constructed portfolio.  While we could add an additional penalty term with a slack variable (similar to what will be described in subsequent sections), a preferable method found in \cite{palmer2021quantum} is to create the mapping:

\begin{equation}
w_n = \frac{1}{P} \left(\sum_{d=0}^{N_d-2}2^d x_{n,d} + Mx_{n,N_d-1} \right )
\end{equation}

where $M = Pw^{max}_{n} - 2^{N_d - 1} - 1$ and $N_d$ is chosen such that $2^{N_d-1} -1\leq Pw^{max}_{n}$.  For example, if $P = 1000$ and $w^{max}_{n} = 2.5\%$, then $N_d = 5$ and M = 10.  This mapping directly imposes an upper limit on the weight of each asset, and has the advantage that there are no unnecessary bit variables included with the caveat that we will have to specify this mapping for different asset-specific bounds.

\subsection{Minimum and Maximum Sector Bands}

The mapping specified above has also been used to create sector bands, i.e. the maximum and minimum proportion of each industry sector we would like our constructed portfolio to have.  One approach to doing this \cite{palmer2021quantum} was to distribute equally the maximum sector bands across the individual assets.  This has a couple disadvantages, one being that by doing so we are forcing the portfolio to include every asset in that sector with a high enough minimum band.  Another disadvantage is that we should allow for the flexibility of allowing a subset of stocks to fill the entire quota for that sector, while also satisfying the separate asset-specific maximum investment amount.  For instance, if we would only like 8\% of our portfolio to be from the healthcare sector while preventing one asset from being more than 2\% of our portfolio, we should still allow for the possibility of only 4 healthcare stocks being selected for that sector.  In our results this formulation allows for more flexibility and more realistic input bands for sectors and assets.

We can therefore introduce more concrete investment bands on sectors by introducing slack variables and creating soft constraints for maximum sector allocation:

\begin{equation}
\sum_{s} \left(\sum_{n} w_{n,s} + s_{s,u} - U_s \right)^2
\end{equation}

and minimum sector allocation:

\begin{equation}
\sum_{s} \left(\sum_{n} w_{n,s} - s_{s,l} - L_s \right)^2
\end{equation}

where $s_s \geq 0$, $U_s$ and $L_s$ is our maximum and minimum percent allocation for that sector, and $w_{n,s}$ is asset $n$ in sector $s$.  Here $s_{s,u}$ and $s_{s,l}$ are binary encoded integer-slack variables.

In Figure 1 we show the results of running the portfolio optimization model on the entire S\&P 500 with certain minimum and maximum investments per sector on a QA.  As compared to results elsewhere, our implementation allowed for much tighter investment bands, more flexibility, and the hybrid solver was able to satisfy all constraints formulated as penalty terms in the objective function.  

\begin{figure}[h]
    \centering
    \includegraphics[width=240px]{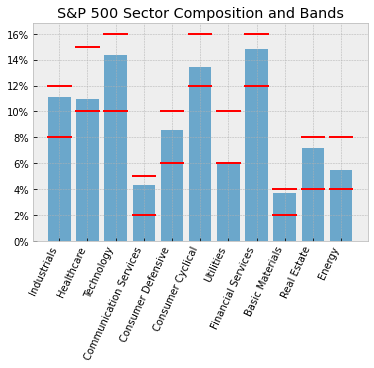}
    \caption{Results for constraining sectors using the Hybrid BQM solver.  The red lines are the minimum and maximum allowable investments per sector and the blue bars are the optimal portfolio composition found.}
    \label{fig:sector_bands}
\end{figure}

\subsection{Balance Sheet Constraints}

In many scenarios, we would like to consider more aspects of an asset rather than their historical returns and volatilities we use to construct our portfolio.  This can arise from the desire to have the portfolio exhibit certain characteristics the investor desires as well as to hedge against biased historical data that we use to formulate our model.  This would be highly beneficial to asset managers who believe in fundamental analysis of stocks and would like their portfolios to represent their belief in various economic indicators.  

An example we used for our portfolios was that of the current ratio, which is a measure of a firm's ability to cover short-term obligations, typically within the next year.  Companies with current ratios greater than 1 can be thought of as more financially solvent than those with ratios less than 1.  Here we show two different realizations of this belief: one where we impose local constraints that each asset should have a minimum current ratio, and another global constraint where our entire portfolio should have a minimum average.  The former would result from the belief that we do not want any stock in our portfolio that does not have a strong asset-liability ratio, while the later would allow for some assets with lower current ratios as long as the entire portfolio meets a minimum level.

The local constraint could be solved by filtering the data before modelling, however its advantage is realized as a soft constraint where the investor has a bias for assets above a certain threshold, but does not want to preclude consideration of investment entirely.  Therefore this constraint is most powerful not as a hard constraint but rather as an extension of an investor's preferences.

The local constraint is:

\begin{equation}
\sum_{n} \left(C_n w_n - s_{n,c} - C_{min} w_n \right)^2
\end{equation}

where $C_n$ is asset $n$'s current ratio, $s_{n,c} \geq 0$ is the binary encoded integer slack variable, and $C_{min}$ is our minimum required ratio.

From this we can also derive the global constraint:

\begin{equation}
\left(C^T w - s_c - C_{min}\right)^2
\end{equation}

where again $s_c \geq 0$.

Our results show how effective both quadratic constraints are when run with appropriate Lagrange Multipliers.  Both constraints offer flexibility to investors depending on their preference for assets being included entirely or for a specific characteristic of the portfolio.  A combination of both these constraints is also feasible, with investors preferring a minimum balance sheet metric for an asset while still wanting the constructed portfolio to have a targeted measure.

\begin{figure*}[t]
    \centering
    \includegraphics[width=240px]{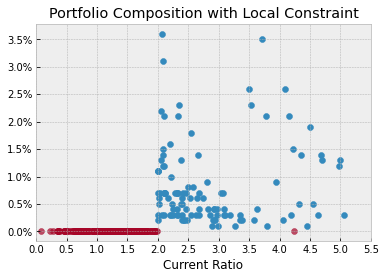}
    \includegraphics[width=240px]{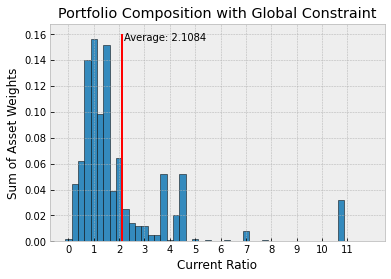}
    \caption{Distribution of current ratio for the local and global constraint on the S\&P 500.  The left graph shows the soft constraint penalized assets whose current ratio was less than 2, the effectiveness of which we can see by the red dots showing those assets were not selected as investments.  The right graph allows for assets with current ratio less than 2, as long as the weighted average of the ratio is greater than 2.}
    \label{fig:cr_combined}
\end{figure*}

\subsection{Cardinality Constraint}

Another constraint we considered is the cardinality constraint found in the Limited Asset Markowitz (LAM) model.  This constraint is significant for several reasons.  The first is that the requirement of limiting the number of assets to select is relevant to portfolio managers and individual investors alike.  Transaction costs, monitoring overhead, and many other considerations typically require an investor to limit the number of assets to choose.  To limit the number of assets requires binary variables, transforming the traditional model into a Mixed Integer Quadratic Programming (MIQP) problem \cite{2014cesarone} which is known to be NP-Hard.

To limit the number of assets to $l$, we can first introduce a binary variable $b_n$ for each stock which is set to 1 if it is invested in and 0 otherwise:

\begin{equation}
\begin{aligned}
&\left(w_n - b_n\right)<=0, \forall { n} \in N \\
& \sum_{n} b_n <= l
\end{aligned}
\end{equation}

To make these constraints amenable to a QUBO problem, we will have to introduce additional binary-encoded integer slack variables $v$ with appropriate Lagrange Multipliers:

\begin{equation}
\begin{aligned}
&(\sum_{n} b_n + v - l)^2 \\
& \sum_{n} (w_n - b_n)
\end{aligned}
\end{equation}

We can see that the first constraint will force the number of binary variables to be less than or equal to the upper limit on the number of assets, and the second forces the binary variable $b_n$ to be 1 if asset $n$ is invested in as this will decrease the cost function.  In this way there may be times that when an asset is not invested in the binary variable will still be 1, however as we are setting an upper constraint on the number of assets, this will suffice to ensure that if an asset is indeed selected, the binary variable will always be 1.

\section{Model Comparisons}

As current QA do not have the qubit capacity to handle our QUBO model directly, we can instead employ the use of D-Wave's hybrid solvers.  These solvers use a mix of classical optimization techniques and the annealers in combination to find feasible solutions.  While the exact methodology of this hybrid approach is proprietary, these solvers open the possibility of solving very large problems while still leveraging the power of current annealers. 

The Binary Quadratic Model (BQM) accepts as input binary variables and an objective function.  It is equivalent to the standard QUBO model and the formulations we have presented thus far were inputted directly into the solver.  The Constrained Quadratic Model (CQM) accepts both integer and binary variables as well as explicit equality and inequality constraints.  This not only makes the constraints we desire easier to handle, but the performance of solver is significant on results run for the full dataset of the S\&P 500.  For the classical optimizer, we employed the use of CPLEX, which is one of the leading Integer and Mixed Integer optimization suites available.

In figure 3 we show the annualized return and the volatility for the portfolios constructed from all models (BQM, CQM, and CPLEX) for different values of $q$.  The only constraints we explicitly stated was that the full budget was used and an asset could not be more than 2.5\% of the total portfolio, however we also ran the CQM model with the volatility as a constraint, which is shown in purple.   We can see that for low values of $q$ (higher volatility), the CQM model significantly outperformed BQM.  This is most likely due to the native support for integers and bounds on those variables.  For higher values of $q$ however, the BQM solver outperformed CQM.  The cutoff appeared to be at $q=25$, where above that coefficient the increased emphasis on the quadratic terms favored the BQM solver which appeared to handle quadratic penalty coefficients better.  We hypothesize this is most likely the result of the use of the annealer in the algorithm, which can natively support binary variables and quadratic terms.  The CQM with target volatility generally underperformed the other models and the results were not as consistent.

\begin{figure}[h!]
    \centering
    \includegraphics[width=240px]{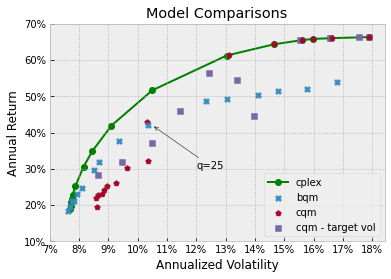}
    \caption{Comparison of results from the BQM and CQM models for different values of q.  We used a range of q values from .1 to 500.  The crossover point found was q = 25, after which the BQM dealt better with the higher values in the quadratic terms and found higher returns for the same volatility as CQM.  Explicitly stating the target volatility as a constraint, showed in purple. did not outperform the standard CQM model.}
    \label{fig:cr_combined}
\end{figure}

In Figure 4 we compare the implementations of all the constraints we have discussed thus far in both the BQM and the CQM with a $q$ value of 1 used for all experiments.  It is evident that the CQM hybrid capabilities at enforcing explicit constraints while minimizing the objective function are superior to BQM for this integer-valued portfolio optimization problem.  Again, the caveat is that with a larger $q$ value or smaller annualized volatility constraint, the BQM solver can potentially outperform CQM.  While we leave it to further research to experiment with the endless combinations of constraints and objectives possible, we posit that the CQM would most likely outperform BQM for most instances.

In the comparison of classical and quantum solutions, we have ran the model with varying levels of constraints and thus difficulty for analyzing the performance of all approaches.  In Figure 3 we can see that for the standard mean-variance portfolio optimization problem, the classical solution found the efficient frontier with minimal effort.  Even with the wide variety of real-world constraints we added to the model in Figure 4, the CPLEX solver outperformed all others in terms of Sharpe Ratios.  

Many have proposed portfolio optimization as a prime candidate for quantum advantage; however the real-world constraints we have discussed thus far show that at least in the static integer-valued case, it is unlikely to outperform classical solutions.  Classical MIQP solvers have been matured for decades and can include millions of variables; it would be surprising that in the vast majority of cases it would underperform others in optimizing portfolios.  Furthermore, many have stated that adding the target volatility as a constraint makes the problem non-convex and thus would allow QA an advantage compared to classical algorithms in the near future.  While this may be true, it is again hard to foresee a real-world scenario in which this is a critical constraint, as it is just as easy to enforce the risk appetite with the $q$ penalty coefficient from Equation 2 in classical solvers as it is required in QA.

\begin{figure}[h]
    \centering
    \includegraphics[width=240px]{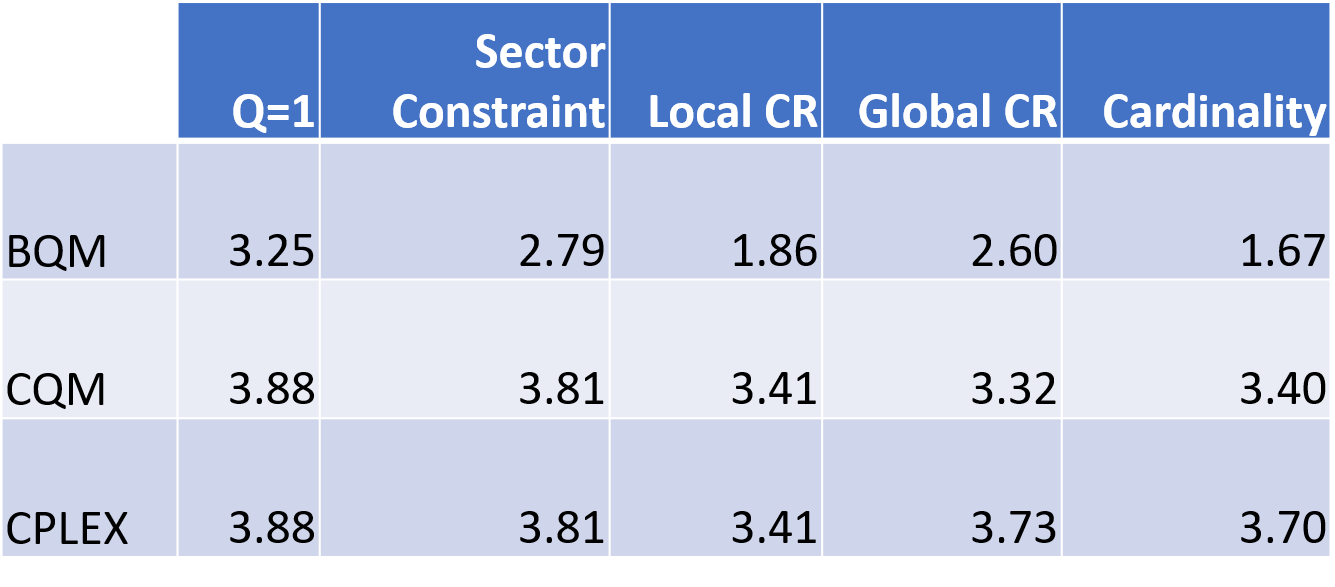}
    \caption{Comparison of Sharpe Ratios for the variety of constraints we have considered across the different models.}
    \label{fig:cr_combined}
\end{figure}

\section{Conclusion}

Here we have shown multiple ways to add flexible and realistic constraints to the portfolio optimization problem.  Importantly, we have formulated them for use in a QUBO problem amenable to quantum annealers.  We demonstrated their effectiveness on the full S\&P 500 dataset using D-Wave's hybrid solvers.  We have also provided direct comparisons for both D-Wave's hybrid solvers and classical solvers, showing the performance of the solvers with current hardware and real-world constaints.

\section{Disclaimer}

About Deloitte: Deloitte refers to one or more of Deloitte Touche Tohmatsu Limited (“DTTL”), its global network of member firms, and their related entities (collectively, the “Deloitte organization”). DTTL (also referred to as “Deloitte Global”) and each of its member firms and related entities are legally separate and independent entities, which cannot obligate or bind each other in respect of third parties. DTTL and each DTTL member firm and related entity is liable only for its own acts and omissions, and not those of each other. DTTL does not provide services to clients. Please see www.deloitte.com/about to learn more.

Deloitte is a leading global provider of audit and assurance, consulting, financial advisory, risk advisory, tax and related services. Our global network of member firms and related entities in more than 150 countries and territories (collectively, the “Deloitte organization”) serves four out of five Fortune Global 500® companies. Learn how Deloitte’s
approximately 330,000 people make an impact that matters at www.deloitte.com. 
This communication contains general information only, and none of Deloitte Touche Tohmatsu Limited (“DTTL”), its global network of member firms or their related entities (collectively, the “Deloitte organization”) is, by means of this communication, rendering professional advice or services. Before making any decision or taking any action that
may affect your finances or your business, you should consult a qualified professional adviser. No representations, warranties or undertakings (express or implied) are given as to the accuracy or completeness of the information in this communication, and none of DTTL, its member firms, related entities, employees or agents shall be liable or
responsible for any loss or damage whatsoever arising directly or indirectly in connection with any person relying on this communication. 
Copyright © 2021. For information contact Deloitte Global.

\bibliographystyle{plain} 
\bibliography{b1.bib}

\begin{thebibliography}{1}

\bibitem{bouland2020prospects}
Adam Bouland, Wim van Dam, Hamed Joorati, Iordanis Kerenidis, and Anupam
  Prakash.
\newblock Prospects and challenges of quantum finance, 2020.

\bibitem{2014cesarone}
Francesco Cesarone, Andrea Scozzari, and Fabio Tardella.
\newblock Linear vs. quadratic portfolio selection models with hard real-world
  constraints.
\newblock {\em Computational Management Science}, 12(3):345–370, May 2014.

\bibitem{cohen2020portfolio}
Jeffrey Cohen, Alex Khan, and Clark Alexander.
\newblock Portfolio optimization of 40 stocks using the dwave quantum annealer,
  2020.

\bibitem{cohen2020portfolio_60}
Jeffrey Cohen, Alex Khan, and Clark Alexander.
\newblock Portfolio optimization of 60 stocks using classical and quantum
  algorithms, 2020.

\bibitem{herman2022survey}
Dylan Herman, Cody Googin, Xiaoyuan Liu, Alexey Galda, Ilya Safro, Yue Sun,
  Marco Pistoia, and Yuri Alexeev.
\newblock A survey of quantum computing for finance, 2022.

\bibitem{markowitz}
Harry Markowitz.
\newblock Portfolio selection.
\newblock {\em The Journal of Finance}, 7(1):77--91, 1952.

\bibitem{hybrid_holding}
Samuel Mugel, Mario Abad, Miguel Bermejo, Javier Sánchez, Enrique Lizaso, and
  Román Orús.
\newblock Hybrid quantum investment optimization with minimal holding period.
\newblock {\em Scientific Reports}, 11(1), Oct 2021.

\bibitem{mugel2020use}
Samuel Mugel, Enrique Lizaso, and Roman Orus.
\newblock Use cases of quantum optimization for finance, 2020.

\bibitem{palmer2021quantum}
Samuel Palmer, Serkan Sahin, Rodrigo Hernandez, Samuel Mugel, and Roman Orus.
\newblock Quantum portfolio optimization with investment bands and target
  volatility, 2021.

\end{thebibliography}

\end{document}